\documentclass[aps,prd,twocolumn,nofootinbib,superscriptaddress,floatfix]{revtex4-2}

\usepackage{amsmath,amssymb,bm}
\usepackage{booktabs}
\usepackage{graphicx}
\usepackage{microtype}
\usepackage[dvipsnames]{xcolor}
\definecolor{valecol}{rgb}{0,0.5, 1.}

\usepackage[colorlinks=true,citecolor=MidnightBlue,linkcolor=MidnightBlue,
          urlcolor=MidnightBlue]{hyperref}
\graphicspath{{figs/}}

\newcommand{\dd}{\mathrm{d}}
\newcommand{\Hperp}{H_{\perp}}
\newcommand{\Hpar}{H_{\parallel}}
\newcommand{\Hloc}{H_{\mathrm{loc}}}
\newcommand{\ThetaS}{\Theta_{\mathrm{M26}}}
\newcommand{\LCDM}{\Lambda\mathrm{CDM}}
\newcommand{\LLTB}{\Lambda\mathrm{LTB}}

\begin{document}

\title{Near universality of nonlinear transverse and radial velocity responses in spherical collapse with arbitrary radial profiles}


\author{Valerio Marra}
\affiliation{Departamento de Física, Universidade Federal de Ouro Preto, 35400-000, Ouro Preto, MG, Brazil}
\affiliation{INAF -- Osservatorio Astronomico di Trieste, via Tiepolo 11,
34131 Trieste, Italy}
\affiliation{IFPU -- Institute for Fundamental Physics of the Universe,
via Beirut 2, 34151 Trieste, Italy}

\date{\today}

\begin{abstract}
The nonlinear relation between density and expansion is usually formulated for
a homogeneous spherical top hat, which has a single local Hubble rate.  A
smooth spherical profile instead expands differently along ($\Hpar$) and across ($\Hperp$) the
radial direction--a direct signature of radial inhomogeneity.  We show that, for growing-mode pressureless matter with a
cosmological constant, the complete shellwise response is nevertheless fixed
by one top-hat function.  Given the local density contrast $\delta(t,r)$ and
the enclosed contrast $\Delta(t,r)$, a transverse response and its derivative determine
$\Hperp$ and $\Hpar$.  The linear and second-order limits are algebraic
maps whose only dynamical input is the usual growth rate~$f$.
An exact equal-age construction supplies the nonlinear response without
integrating an evolution equation.  We check that it reconstructs full $\LLTB$ profiles to
numerical precision.  We also
provide a derivative-aware, cosmology-independent three-term symbolic fit that requires only $f$, $\delta$, and $\Delta$.  Across a representative set of matter-curvature-redshift combinations and for a shell located in the compensated transition, the maximum relative errors are $0.3\%$ and $0.7\%$ in the transverse and radial responses, respectively.  This compact formulation separates
the production of a density profile from its expansion response and makes the
effect of radial gradients explicit.
\end{abstract}

\maketitle

\section{Introduction}
\label{sec:introduction}

The density and peculiar-velocity fields provide complementary views of
structure formation.  The density field describes where matter accumulates,
while the velocity field records how it is still evolving.  In linear theory
their relation is controlled by the logarithmic growth rate
$f=\dd\ln D/\dd\ln a$.
{Here $D(a)$ is the linear growing-mode matter density growth factor
of the reference FLRW background; its constant normalization does not affect~$f$.}  Nonlinearly, the relation is generally complicated by
the shape and environment of a perturbation.  Spherical collapse provides the
cleanest setting in which it can be studied beyond linear theory while
retaining an exact dynamical description.

The nonlinear spherical density--velocity relation has a long history.
Bernardeau obtained a remarkably compact result in the zero-variance limit
and connected spherical collapse to the perturbative hierarchy of density and
velocity statistics
\cite{1992ApJ...390L..61B,1992ApJ...392....1B}.  Subsequent work quantified
the effects of finite variance and smoothing \cite{Chodorowski:1996bu}, and
showed that spherical collapse continues to organize the perturbative result
for general matter densities, with only a weak residual cosmology dependence
after the leading growth-rate scaling is removed \cite{Fosalba:1998da}.
Accurate analytic relations were later constructed for both voids and
overdensities \cite{Bilicki:2008rx}.  Nadkarni-Ghosh extended this picture to
flat constant-$w$ cosmologies by selecting the common-age growing solution in
density--velocity phase space \cite{Nadkarni-Ghosh:2012hmw}.  More recently,
joint density and velocity statistics in concentric spheres have shown how
useful the same relation remains in the mildly nonlinear regime
\cite{Uhlemann:2016wug}.
In a study of cosmic variance in the local Hubble constant, Marra
et al.\ wrote the expansion perturbation of a homogeneous spherical
region as the product of its linear prediction and a nonlinear correction~$\Theta$. They provided an accurate fit for $\Theta$ and found, in agreement
with earlier work, only a weak dependence on the background cosmology
\cite{Marra:2013rba}. This result suggests that much of the nonlinear
spherical response can be captured by a simple, nearly universal relation.

These classic relations describe a homogeneous top hat, which has only one
expansion rate.  A smooth spherical profile contains more information.  The
transverse area of a matter shell and the radial distance between neighboring
shells need not grow at the same rate.  Their difference is generated by the
radial variation of the profile and vanishes in a homogeneous interior.
Consequently, extending a top-hat relation to an arbitrary radial profile
requires a prediction for both the transverse and radial velocity responses.
The transverse part of this extension was already identified by
Nadkarni-Ghosh, who showed that the top-hat density--velocity relation can be
applied shell by shell to an arbitrary spherical infall profile when written
in terms of the enclosed contrast~$\Delta$
\cite{Nadkarni-Ghosh:2012hmw}.  The additional step developed here is to
determine the radial and local expansion rates and anisotropy~$\Gamma$.  These depend on
the local-to-enclosed difference $\delta-\Delta$ and on the derivative of the
transverse response.

Exact spherical dynamics already provides the framework for this extension.
The Lema{\^\i}tre--Tolman--Bondi (LTB) solution describes general pressureless
spherical profiles \cite{Lemaitre1933,Tolman1934,Bondi1947}.  Sussman's
quasi-local formulation makes the distinction between the density inside a
shell and the density at that shell explicit, and relates the corresponding
expansion rates through exact radial identities
\cite{Sussman:2008wx,Sussman:2012xc,Sussman:2013yq}.  Its growing-mode and
linear limits have also been characterized
\cite{Sussman:2013qya,Sussman:2014wua}.  The full $\LLTB$ evolution and its
connection to relativistic and Newtonian descriptions were reviewed in Ref.~\cite{Marra:2022ixf};
exact analytic representations were
developed in Ref.~\cite{Valkenburg:2011tm}.  Our purpose is not to introduce a
new spherical solution, but to turn this structure into a compact response
written in directly interpretable density variables.

We show that one function $A(\Delta)$ and its ordinary derivative
$A'(\Delta)$ determine the transverse, radial, and local expansion rates of
an arbitrary smooth spherical profile before shell crossing.  The result
extends the familiar top-hat response without requiring a new fit for each
profile.  Conceptually, it separates the evolution that produces the density
profile from the response studied here:
\begin{equation}
\begin{aligned}
k(r)&\xrightarrow{\text{exact evolution}}
\bigl\{\delta(t,r),\Delta(t,r)\bigr\}\\
&\xrightarrow{\text{shell response}}
\bigl\{\Hperp,\Hpar,\Hloc,\Gamma\bigr\}.
\end{aligned}
\label{eq:conceptual-map}
\end{equation}
The first arrow determines how a chosen curvature profile evolves; we refer
to Ref.~\cite{Marra:2022ixf} for that calculation.  The present work concerns the second
arrow.  It can therefore also be used when the density profiles come from a
different spherical solver or from a phenomenological model.  We derive the
exact common-age response and its linear and second-order expansions, and
construct a fast, derivative-aware nonlinear approximation that can be
applied shell by shell.

We first formulate the geometry and the two required density contrasts in
Sec.~\ref{sec:geometry}.  Section~\ref{sec:response} derives the exact
shellwise response and its linear and second-order limits.  The fast nonlinear approximation is
introduced in Section~\ref{sec:closures}.  Section~\ref{sec:results} tests the
construction across cosmologies and against complete $\LLTB$ profiles.  We
discuss the interpretation, scope, and conclusions in
Section~\ref{sec:discussion}.  Appendix~\ref{app:calibration} documents the
re-optimization of the universal fit.
{Appendix~\ref{app:covariant} connects our variables to covariant
kinematics and the established $q$-scalar evolution equations.}
Quantities without an explicit $r$
argument refer to the FLRW background; $r$ is a Lagrangian label attached to
the matter flow; and $c=1$ throughout.

\section{Spherical geometry and density variables}
\label{sec:geometry}

\subsection{Directional expansion}

We use the notation of Ref.~\cite{Marra:2022ixf}.  The metric of a pressureless, spherically
symmetric spacetime with a cosmological constant is
\begin{equation}
\dd s^2=-\dd t^2
+\frac{a_\parallel^2(t,r)}{1-k(r)r^2}\dd r^2
+a_\perp^2(t,r)r^2\dd\Omega^2,
\label{eq:metric}
\end{equation}
where
\begin{equation}
R(t,r)\equiv r a_\perp(t,r),
\qquad
a_\parallel(t,r)=R'(t,r).
\label{eq:scale-factors}
\end{equation}
A prime and a dot denote partial derivatives with respect to $r$ and $t$,
respectively.  The coordinate $r$ labels the same dust shell at all times; it
is therefore a Lagrangian matter coordinate rather than an instantaneous
physical distance.

The transverse and radial expansion rates are
\begin{equation}
\Hperp(t,r)=\frac{\dot R}{R}
=\frac{\dot a_\perp}{a_\perp},
\qquad
\Hpar(t,r)=\frac{\dot R'}{R'}.
\label{eq:directional-H}
\end{equation}
{These rates are the tangential and radial eigenvalues of the expansion
tensor of the dust flow, giving them a coordinate-independent interpretation.
Their relation to the expansion scalar and shear is given in
Appendix~\ref{app:covariant-kinematics}.}
Their local volume-expansion rate and a dimensionless anisotropy are
\begin{equation}
\Hloc(t,r)=\frac{2\Hperp+\Hpar}{3},
\qquad
\Gamma(t,r)=\frac{\Hpar-\Hperp}{\Hloc}.
\label{eq:Hloc-Gamma}
\end{equation}
The FLRW limit is $a_\perp=a_\parallel=a(t)$, for which both rates equal
$H(t)$ and $\Gamma=0$.  Regularity also gives $\Gamma(t,0)=0$, but a smooth
inhomogeneous profile generally has $\Gamma\ne0$ away from its center.

\subsection{Local and enclosed density contrasts}

\begin{figure}
\centering
\includegraphics[width=\columnwidth]{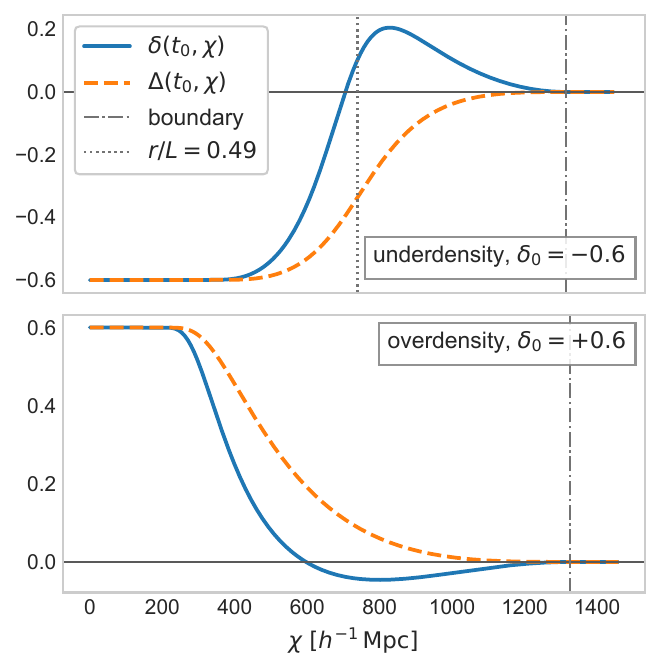}
\caption{Present-day local $\delta$ and enclosed $\Delta$ for compensated
profiles with $\delta_0=-0.6$ (top) and $+0.6$ (bottom), versus the Eulerian
comoving radius $\chi$. Dash-dotted lines mark the matching radii; the dotted line in
the upper panel marks the representative shell $r/L=0.49$ used in
Section~\ref{sec:across-bkg}.}
\label{fig:density-profiles}
\end{figure}

Two contrasts are necessary.  The local one is
\begin{equation}
\delta(t,r)=\frac{\rho_m(t,r)}{\rho_m(t)}-1,
\label{eq:local-delta}
\end{equation}
where $\rho_m(t)$ is the matter density of the reference FLRW background.
{The fields $\rho_m(t,r)$ and $\rho_m(t)$ are distinct: the former is the
local dust density, whereas the latter is the homogeneous reference density
at the same proper time.  For the compensated examples below, this reference
is fixed by the matched FLRW exterior.  Within our notation
local quantities are distinguished by their explicit $r$ dependence; in Appendix~\ref{app:covariant}
we write $\rho_b(t)\equiv\rho_m(t)$ and $H_b(t)\equiv H(t)$ to make the
background identification explicit.}
If $M(r)$ is the conserved matter mass inside shell $r$, the enclosed contrast is
\begin{equation}
\Delta(t,r)
=\frac{3M(r)}{4\pi\rho_m(t)R^3(t,r)}-1.
\label{eq:enclosed-Delta}
\end{equation}
{Because the reference density $\rho_m(t)$ is spatially homogeneous,
$\Delta$ is also the average of the local contrast $\delta$ inside the
shell. In Sussman's $q$-scalar notation,
$\Delta=\langle\delta\rangle_q$, with enclosed density
$\rho_q=3M/(4\pi R^3)$.
This average uses the Euclidean-volume measure of Eq.~(22) of
Ref.~\cite{Marra:2022ixf}, not as an approximation but as the exact
quasi-local volume associated with the Misner--Sharp mass in spherical
symmetry \cite{Sussman:2008wx,Sussman:2012xc,Sussman:2013yq}.
Appendix~\ref{app:q-dictionary} specifies the averaging measure and
relates our contrasts to the $q$-scalar fluctuations.}

The contrast $\delta$ answers a local question, whereas $\Delta$ measures the
mean mass excess or deficit enclosed by the shell.  They coincide in the
homogeneous interior of a top hat and at a regular symmetry center.  In a
compensated profile they can have opposite signs: a shell in the overdense
wall of a void can be locally overdense while still enclosing a net mass
deficit.  Figure~\ref{fig:density-profiles} illustrates this distinction for
the two smooth compensated profiles used below. We plot against the shell's Eulerian comoving radius,
defined relative to the reference FLRW background as $\chi(t,r)\equiv R(t,r)/a(t)$,
rather than against its Lagrangian label~$r$, which is subject to gauge choices.  The local contrast changes sign in the
compensating shell, while the enclosed contrast approaches zero at the
matching boundary.  This distinction will control the directional response.

\section{The shellwise response}
\label{sec:response}

\subsection{Exact transverse closure}

We normalize the Lagrangian shell label in the early FLRW regime.  Mass
conservation then gives \cite{Marra:2022ixf}
\begin{equation}
a_\perp(t,r)=a(t)\,[1+\Delta(t,r)]^{-1/3}.
\label{eq:mass-conservation}
\end{equation}
At fixed Lagrangian shell $r$, the LTB energy equation takes the form of a
Friedmann equation. For later use, it is convenient to write it as
\begin{equation}
\frac{\Hperp^2(t,r)}{H_0^2}
=\Omega_{m0}a_\perp^{-3}
+\widehat{\Omega}_{k0}(r)a_\perp^{-2}
+\Omega_{\Lambda0},
\label{eq:shell-friedmann}
\end{equation}
while the background obeys
\begin{equation}
E^2(a)\equiv\frac{H^2(a)}{H_0^2}
=\Omega_{m0}a^{-3}
+\Omega_{k0}a^{-2}
+\Omega_{\Lambda0}.
\label{eq:background-friedmann}
\end{equation}
Here
$\widehat{\Omega}_{k0}(r)\equiv-k(r)/H_0^2$
is normalized by the square of the present-day background Hubble rate and is distinct from
the shell-normalized $\Omega_{k0}(r)$ of Ref.~\cite{Marra:2022ixf}. In the
matched FLRW region,
$\widehat{\Omega}_{k0}(r)=\Omega_{k0}$.
We assume
$\Omega_{\Lambda0}=1-\Omega_{m0}-\Omega_{k0}$
and neglect radiation.

Equation~\eqref{eq:mass-conservation} fixes $a_\perp$ once the instantaneous
$\Delta$ and the background epoch $a$ are given. The growing mode is selected
by requiring the shell and background to have the same age,
\begin{align}
H_0t(a)
&\equiv \int_0^a\frac{\dd\tilde a}{\tilde a E(\tilde a)}
\nonumber\\
&=\int_0^{a_\perp(t,r)}
\frac{\dd\tilde a}{\tilde a
\sqrt{\Omega_{m0}\tilde a^{-3}
+\widehat{\Omega}_{k0}(r)\tilde a^{-2}+\Omega_{\Lambda0}}}
\nonumber\\
&\equiv {\cal T}\!\left[a_\perp(t,r),
\widehat{\Omega}_{k0}(r)\right] \,.
\label{eq:equal-age}
\end{align}
Here ${\cal T}$ denotes the dimensionless shell-age functional; the equal-age
condition sets it equal to the dimensionless background age $H_0t(a)$.
This one-dimensional relation determines $\widehat{\Omega}_{k0}(r)$ on the
expanding branch.
It is the familiar separate-universe construction of spherical collapse,
expressed for an arbitrary Lagrangian shell
\cite{Nadkarni-Ghosh:2012hmw,Dai:2015jaa}.  In LTB language, the simultaneous
bang time removes the decaying density mode \cite{Zibin:2008vj}.  No
time-evolution differential equation is solved.

To separate the linear normalization from the nonlinear response,
define
\begin{equation}
\delta_{h_\perp}=-\frac{f}{3}A(\Delta),
\qquad
A(\Delta)=\Delta \cdot \Theta_\perp(\Delta),
\label{eq:A-definition}
\end{equation}
{where $\delta_{h_i}(t,r)=H_i(t,r)/H(t)-1$, with
$i\in\{\perp,\parallel,\mathrm{loc}\}$.  The denominator is always the
reference FLRW rate $H(t)=\dot a(t)/a(t)$, evaluated at the same proper time
as the shell quantities.
Thus $A$ is the enclosed contrast multiplied by its nonlinear correction
$\Theta_\perp$. In the linear limit, $\Theta_\perp\to1$ and
$A(\Delta)\simeq\Delta$.}
In the exact construction, the response is more fully
$A(\Delta;a,\Omega_{m0},\Omega_{k0})$; throughout, its fixed background and
epoch arguments are suppressed to emphasize the shell-contrast dependence.
For matter, curvature, and a cosmological constant, the exact growing mode
can be written with the Heath integral
\cite{1977MNRAS.179..351H}
\begin{equation}
D(a)\propto E(a)\int_0^a
\frac{\dd \tilde a}{\tilde a^3E^3(\tilde a)}.
\label{eq:heath}
\end{equation}
For flat $\LCDM$, this is equivalently a hypergeometric function \cite{Silveira:1994yq}.  The
nonlinear function $A$ is obtained from
Eqs.~\eqref{eq:mass-conservation}--\eqref{eq:A-definition}.

The derivative $A'\equiv\partial A/\partial\Delta$ is needed below and can be
obtained without finite differences. Implicit differentiation of
Eq.~\eqref{eq:equal-age} gives
\begin{equation}
\frac{\dd\widehat{\Omega}_{k0}}{\dd\Delta}
=-\frac{{\cal T}_{,a_\perp}}
{{\cal T}_{,\widehat{\Omega}_{k0}}}
\frac{\dd a_\perp}{\dd\Delta},
\qquad
\frac{\dd a_\perp}{\dd\Delta}
=-\frac{a_\perp}{3(1+\Delta)}.
\label{eq:curvature-coefficient-derivative}
\end{equation}
Differentiating Eq.~\eqref{eq:shell-friedmann} then gives $A'$ analytically.
In particular, $A(0)=0$ and $A'(0)=1$.

\subsection{From a top hat to a general shell}
\label{sec:general-shell}

The central simplification follows from two exact identities.  First,
Eq.~\eqref{eq:scale-factors} implies
\begin{equation}
\Hpar=\Hperp+\frac{R}{R'}\partial_r\Hperp.
\label{eq:H-radial-identity}
\end{equation}
Second, the enclosed mass obeys $M'=4\pi\rho_m(t,r)R^2R'$.
Differentiating Eq.~\eqref{eq:enclosed-Delta} therefore gives
\begin{equation}
\frac{R}{R'}\partial_r\Delta=3(\delta-\Delta).
\label{eq:Delta-radial-identity}
\end{equation}
This is the density specialization of the exact $q$-scalar gradient identity
\cite{Sussman:2008wx,Sussman:2012xc,Sussman:2013yq}.
At a fixed time, $H$ and $f$ have no $r$ dependence, while the shell response
depends on position through $\Delta(t,r)$.
Applying the chain rule to
Eq.~\eqref{eq:A-definition} therefore yields
\begin{equation}
\delta_{h_\parallel}
=-\frac{f}{3}A(\Delta)
-f[\delta-\Delta]A'(\Delta).
\label{eq:Hparallel-response}
\end{equation}
Combining the directional rates gives
\begin{equation}
\delta_{h_{\mathrm{loc}}}
=-\frac{f}{3}\left\{A(\Delta)
+[\delta-\Delta]A'(\Delta)\right\},
\label{eq:Hlocal-response}
\end{equation}
and
\begin{equation}
\Gamma=
\frac{f[\Delta-\delta]A'(\Delta)}
{1-\frac{f}{3}\left\{A(\Delta)
+[\delta-\Delta]A'(\Delta)\right\}}.
\label{eq:Gamma-response}
\end{equation}

Equations~\eqref{eq:A-definition} and
\eqref{eq:Hparallel-response}--\eqref{eq:Gamma-response} are the main result.
They make the physics transparent.  The transverse rate responds to the mass
enclosed by the shell.  The radial rate additionally responds to the mismatch
between local and enclosed density.  The latter is a direct measure of the
profile gradient.  For a top hat $\delta=\Delta$, so the derivative term
vanishes, $\Hpar=\Hperp$, and $\Gamma=0$ even nonlinearly.

The $q$-scalar identity in Eq.~\eqref{eq:Delta-radial-identity} supplies the
established relation between the radial gradient of the enclosed contrast and
the local--enclosed difference.  Our new step is to combine it with the
equal-age growing-mode closure.  At a fixed background epoch, $fA(\Delta)$
fixes the transverse response, while $f[\delta-\Delta]A'(\Delta)$ is the only
additional term entering the radial, local, and anisotropic responses.  Thus,
once $\delta(t,r)$, $\Delta(t,r)$, the background $H$ and $f$, and the single
response $A$ are specified, all four expansion observables follow without
independent shell-curvature or bang-time data and without numerically
differentiating the metric.
{The underlying geometric identities are established results of the
$q$-scalar framework.  Appendix~\ref{app:q-dictionary} shows explicitly that
$\delta-\Delta$ measures electric Weyl curvature and that the directional
rate difference measures shear.  The present growing-mode reduction and the
derivative-aware nonlinear fit use these identities to construct the
density--expansion response.}

\subsection{Linear limit}
\label{sec:linear}

For $|\delta|,|\Delta|\ll1$,
\begin{equation}
A(\Delta)=\Delta+{\cal O}(\Delta^2),
\qquad A'(\Delta)=1+{\cal O}(\Delta),
\end{equation}
so that the response becomes
\begin{align}
\delta_{h_\perp}^{(1)}&=-\frac{f\Delta}{3},
\label{eq:linear-perp}\\
\delta_{h_\parallel}^{(1)}&=-f\delta+\frac{2f\Delta}{3},
\label{eq:linear-parallel}\\
\delta_{h_{\mathrm{loc}}}^{(1)}&=-\frac{f\delta}{3},
\label{eq:linear-local}\\
\Gamma^{(1)}&=f(\Delta-\delta).
\label{eq:linear-gamma}
\end{align}
The first-order response is thus completely algebraic.  Its only dynamical
input is $f$; neither $H_0$, an age integral, nor a curvature reconstruction is
required.

\subsection{Second-order expansion}
\label{sec:second-order}

The same reduction remains algebraic at second order.  Let
$\delta,\Delta={\cal O}(\epsilon)$ and expand the spherical response as
\begin{equation}
A(\Delta)=\Delta+c_2\Delta^2+{\cal O}(\epsilon^3).
\label{eq:second-order-A}
\end{equation}
For Einstein--de Sitter spherical collapse, the nonlinear and linearly
extrapolated enclosed contrasts obey
\begin{equation}
\Delta=\Delta_{\rm L}+\frac{\nu_2}{2}\Delta_{\rm L}^2
+{\cal O}(\Delta_{\rm L}^3),
\qquad \nu_2=\frac{34}{21}.
\label{eq:eds-nu2}
\end{equation}
The continuity equation then gives
\begin{equation}
c_2=\frac{\nu_2}{2}-1=-\frac{4}{21}.
\label{eq:eds-c2}
\end{equation}
Equations~\eqref{eq:eds-nu2} and \eqref{eq:eds-c2} are the standard EdS result
for spherical collapse~\cite{1992ApJ...392....1B,Fosalba:1998da}.  We combine
these coefficients with the exact linear growth rate $f$ of the selected
background.  Expanding
Eqs.~\eqref{eq:A-definition} and
\eqref{eq:Hparallel-response}--\eqref{eq:Gamma-response} consistently through
${\cal O}(\epsilon^2)$ yields
\begin{align}
\delta_{h_\perp}^{(2)}={}&-\frac{f\Delta}{3}
-\frac{fc_2\Delta^2}{3},
\label{eq:second-order-perp}\\
\delta_{h_\parallel}^{(2)}={}&-f\delta+\frac{2f\Delta}{3}
-2fc_2\delta\Delta+\frac{5fc_2\Delta^2}{3},
\label{eq:second-order-parallel}\\
\delta_{h_{\rm loc}}^{(2)}={}&-\frac{f\delta}{3}
-\frac{fc_2}{3}\left(2\delta\Delta-\Delta^2\right),
\label{eq:second-order-local}\\
\Gamma^{(2)}={}&f(\Delta-\delta)
\left(1+2c_2\Delta+\frac{f\delta}{3}\right).
\label{eq:second-order-gamma}
\end{align}
Here $\Gamma^{(2)}$ is the strict expansion of
Eq.~\eqref{eq:Gamma-response} through ${\cal O}(\epsilon^2)$, not an
unexpanded ratio of truncated rates.  No equal-age root is required.  This
provides a controlled perturbative benchmark between the linear response and
the finite-contrast closures below; it is not intended to replace them deep
in the nonlinear regime.  For a top hat, $\delta=\Delta$, the two directional
rates remain equal and $\Gamma^{(2)}=0$.

\section{Fast nonlinear closures}
\label{sec:closures}

\subsection{Earlier reference relations}

The scaled velocity divergence used in the classic spherical literature maps
to our function $A=\Delta\Theta_\perp$, not to the multiplicative correction
$\Theta_\perp$ itself.  In this notation the Bernardeau relation
\cite{1992ApJ...390L..61B} is
\begin{equation}
A_{\rm B92}(\Delta)=\frac{3}{2}
\left[(1+\Delta)^{2/3}-1\right].
\label{eq:B92}
\end{equation}
Bilicki and Chodorowski proposed the
overdensity approximation ($\Delta\geq0$)
\begin{equation}
A_{\rm BC08}(\Delta)=3\left[(1+\Delta)^{1/2}
-(1+\Delta)^{1/6}\right],
\label{eq:BC08}
\end{equation}
which remains accurate into the mildly nonlinear regime and up to spherical
turnaround \cite{Bilicki:2008rx}.  Both have the same second-order expansion,
\begin{equation}
A(\Delta)=\Delta-\frac{1}{6}\Delta^2+{\cal O}(\Delta^3),
\label{eq:classic-expansion}
\end{equation}
This quadratic coefficient is close to, but not identical to, the exact EdS
value $-4/21$ in Eq.~\eqref{eq:eds-c2}; B92 and BC08 are finite-contrast
approximations rather than perturbative truncations.  Moreover,
Eq.~\eqref{eq:BC08} was neither derived nor calibrated for voids.  BC08
give a separate void relation that retains an explicit cosmology dependence;
we therefore do not extrapolate Eq.~\eqref{eq:BC08} to $\Delta<0$.

Nadkarni-Ghosh generalized these relations to flat backgrounds containing
matter and nonclustering dark energy with constant equation of state $w$
\cite{Nadkarni-Ghosh:2012hmw}.  With the instantaneous matter fraction
$\Omega_m(a)$, define
\begin{align}
\gamma_2(w)&=-0.01(-w)^{-1.18},\nonumber\\
B(\Omega_m,w)&=\frac{2}{3}\Omega_m(a)^{\gamma_2},\nonumber\\
C(\Omega_m,w)&=\frac{3}{2}\Omega_m(a)^{-\gamma_2}.
\label{eq:NG13-shape-parameters}
\end{align}
In our convention the resulting
shape response is
\begin{equation}
A_{\rm NG13}(\Delta)=
\begin{cases}
C\left[(1+\Delta)^B-1\right],
&-1<\Delta\leq1,\\[2mm]
A_{\rm BC08}(\Delta),
&1<\Delta\leq10.
\end{cases}
\label{eq:NG13}
\end{equation}
The low-density branch reduces to B92 for $\Omega_m=1$, while the high-density
branch is BC08 after the fitted growth rate is factored out.  NG13 report a
maximum velocity-relation error of about $3\%$ over
$-3/2\leq w\leq-1/2$ and $-1\leq\Delta\leq10$.  The two independently fitted
branches meet only approximately at $\Delta=1$, and no derivative-matching
condition was imposed.

Equations~\eqref{eq:B92}, \eqref{eq:BC08}, and \eqref{eq:NG13} are valuable
analytic benchmarks.
They were designed for the top-hat density--velocity relation, however, and
not optimized for $A'$, which controls the directional response of a smooth
shell.

\subsection{Derivative-aware fit for smooth spherical profiles}

\begin{figure}
\centering
\includegraphics[width=\columnwidth]{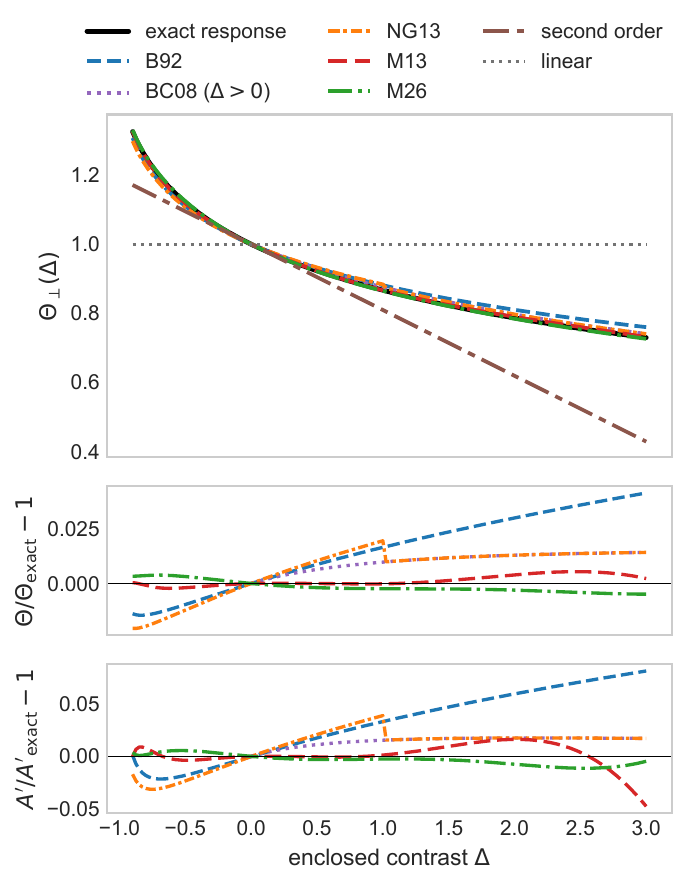}
\caption{Exact nonlinear transverse correction, together with the linear,
second-order, B92, BC08, NG13, M13, and M26
relations.  BC08 is shown only for overdensities, its stated domain; NG13
uses its published $w=-1$ piecewise form.  The lower panels show fractional
residuals in $\Theta_\perp$ and $A'$.  Their vertical ranges resolve the
finite-contrast closures; the second-order residual leaves those ranges away
from $\Delta=0$.}
\label{fig:theta-fiducial}
\end{figure}

The equal-age construction is inexpensive, but repeated curvature roots can
still be undesirable in forecasts or likelihood calculations.  Marra et
al.\ introduced the fit
\begin{equation}
\Theta_{\mathrm{M13}}(\Delta)
=1-0.0882\Delta-
\frac{0.123\sin\Delta}{1.29+\Delta},
\label{eq:M13}
\end{equation}
calibrated at $z=0$ for a flat background with
$\Omega_{m0}=0.3086$ \cite{Marra:2013rba}.  It reproduces the top-hat response
to about $0.4\%$ over the nonlinear interval used there.

For a smooth profile, accuracy in $\Theta$ alone is insufficient because
$\Hpar$, $\Hloc$, and $\Gamma$ depend on
\begin{equation}
A'(\Delta)=\Theta(\Delta)+\Delta\Theta'(\Delta).
\label{eq:Aprime-theta}
\end{equation}
We therefore perform a constrained symbolic search using both $\Theta$ and
$A'$ on a grid of cosmologies and epochs.  The selected three-term relation,
called M26 below, is
\begin{equation}
\begin{aligned}
\ThetaS(\Delta)&=1+0.1320\Delta\\
&-0.6717\left(\sqrt{1+\Delta}-1\right)
+0.013\sin\Delta \,.
\end{aligned}
\label{eq:M26}
\end{equation}
Details of the calibration are given in Appendix~\ref{app:calibration}.
The expression has no fitted denominator, is regular throughout the physical
domain $\Delta>-1$, and satisfies $\Theta(0)=A'(0)=1$ exactly.
Consequently
Eqs.~\eqref{eq:A-definition} and
\eqref{eq:Hparallel-response}--\eqref{eq:Gamma-response} become a root-free
nonlinear response, requiring only the growth rate $f$ and the density
contrasts $\Delta$ and $\delta$, similarly to the linear and second-order results.  The fit has no explicit
cosmological parameter because the nonlinear correction is nearly universal,
as discussed below.  It is
intended for
\begin{equation}
\begin{gathered}
-0.9\le\Delta\le3,\qquad
0.2\le\Omega_{m0}\le0.4,\\
-0.1\le\Omega_{k0}\le0.1,\qquad
0\le z\le5.
\end{gathered}
\label{eq:calibration-domain}
\end{equation}
The exact response should be used outside this domain or when subpercent
control of the residual cosmology dependence is required.

Figure~\ref{fig:theta-fiducial} compares the exact response, the linear and
second-order truncations,
the B92, BC08, and NG13 relations, and the M13 and M26 fits at the
fiducial cosmology of Ref.~\cite{Marra:2013rba}.
Table~\ref{tab:reference-errors} quantifies both
$\Theta_\perp$ and $A'$.  B92 is already accurate at the few-percent level,
and BC08 improves its overdensity branch.  NG13 stays within $2.1\%$ in
$\Theta_\perp$ on both branches, consistent with its published $3\%$
velocity benchmark, but reaches $3.9\%$ in $A'$.  The latter is a new
diagnostic, not a failure of the stated NG13 fit: its two branches were not
optimized or joined for derivative accuracy.  On the void branch M13 remains
slightly better in $\Theta_\perp$, with maximum error $0.2\%$ rather than
$0.4\%$, while M26 improves the corresponding $A'$ error from $0.9\%$
to $0.6\%$.  Over $0<\Delta\le3$, M26 is better in both quantities, with
maximum errors $0.5\%$ in $\Theta_\perp$ and $1.1\%$ in $A'$.  This is
expected: M26 is not optimized for one curve, but for multiple cosmologies,
an extended contrast interval, and the derivative needed by general shells.

\begin{table}
\caption{Maximum absolute fractional error (percent) at the fiducial flat
background, evaluated separately on $-0.9\leq\Delta<0$ and
$0<\Delta\leq3$.  BC08 Eq.~\eqref{eq:BC08} is not a void closure.}
\label{tab:reference-errors}
\begin{ruledtabular}
\begin{tabular}{llrr}
branch & relation & $\Theta_\perp$ & $A'$ \\
\hline
underdensity & B92 & 1.5 & 2.2 \\
     & NG13 & 2.1 & 3.1 \\
     & M13 & 0.2 & 0.9 \\
     & M26 & 0.4 & 0.6 \\
\hline
overdensity & B92 & 4.1 & 8.1 \\
     & BC08 & 1.4 & 1.8 \\
     & NG13 & 2.0 & 3.9 \\
     & M13 & 0.5 & 4.8 \\
     & M26 & 0.5 & 1.1 \\
\end{tabular}
\end{ruledtabular}
\end{table}

\section{Validation and performance}
\label{sec:results}

\subsection{Dependence on cosmology and epoch}

\begin{figure}
\includegraphics[width=\columnwidth]{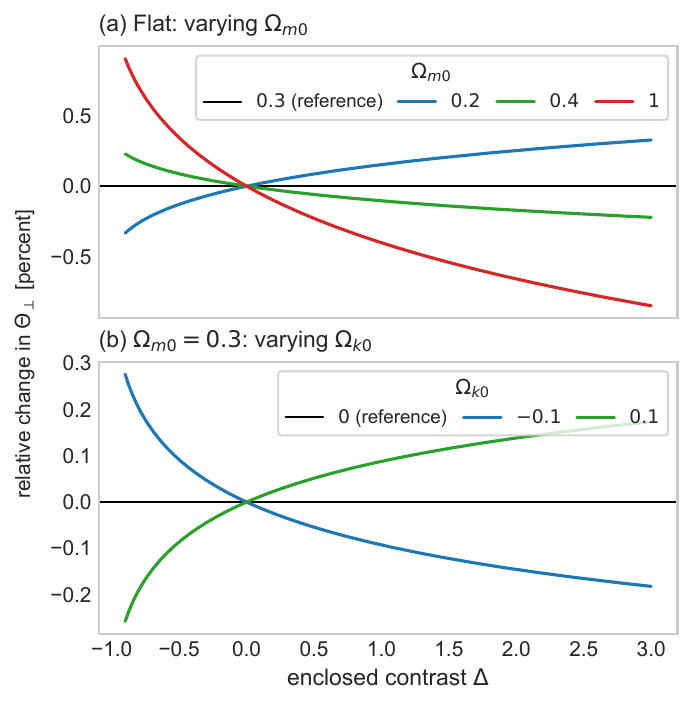}
\caption{Fractional change in the exact $\Theta_\perp(\Delta)$ relative to
$(\Omega_{m0},\Omega_{k0})=(0.3,0)$.  The upper panel varies the present
matter fraction in flat backgrounds; the lower panel varies the background
curvature at fixed $\Omega_{m0}=0.3$.}
\label{fig:theta-cosmology}
\end{figure}

Figure~\ref{fig:theta-cosmology} confirms the classic observation that the
nonlinear density--expansion correction is only weakly cosmology dependent
after the growth rate $f$ has been factored out
\cite{Nusser:1997ec,Bilicki:2008rx,Nadkarni-Ghosh:2012hmw,Marra:2013rba}.
Across the ranges $0.2\leq\Omega_{m0}\leq0.4$ and
$-0.1\leq\Omega_{k0}\leq0.1$ shown here, the fractional change in
$\Theta_\perp$ is at most $0.34\%$, less than the maximum error in any of the
approximate fits of Section~\ref{sec:closures}.  Neglecting the cosmology
dependence is therefore justified in many applications.

The dependence
is nevertheless nonzero.  Varying $\Omega_{m0}$ or $\Omega_{k0}$ changes both
the equal-age curvature and the growth history.  Redshift has the same
status: holding $\Delta$ fixed while changing $z$ probes the response law, not
the trajectory of one physical shell, whose contrasts also evolve.
At fixed density parameters, all dimensionless equations depend on $H_0$ only
through $H_0t$.  It follows that $\Theta_\perp$, $A'$, and $\Gamma$ are
independent of the numerical value of $H_0$, while every dimensional expansion
rate scales linearly with it.  Changing $H_0$ at fixed physical density
$\omega_m=\Omega_{m0}h^2$ is different because it changes $\Omega_{m0}$.

\begin{table}
\caption{Maximum absolute relative error (percent) across the eight cases of Section~\ref{sec:across-bkg}.  The denominators are the exact
dimensionless responses, not the full rates.}
\label{tab:backend-errors}
\begin{ruledtabular}
\begin{tabular}{lcccc}
closure & $\delta_{h_\perp}$ & $\delta_{h_\parallel}$ & $\delta_{h_{\rm loc}}$ & $\Gamma$ \\
\hline
linear & 7.0 & 16.5 & 31.6 & 18.5 \\
second order & 1.1 & 3.9 & 8.5 & 5.3 \\
M13    & 0.32 & 0.81 & 1.60 & 0.78 \\
M26    & 0.30 & 0.70 & 1.34 & 0.63 \\
\end{tabular}
\end{ruledtabular}
\end{table}

\subsection{Approximate closures across backgrounds}
\label{sec:across-bkg}

To activate the derivative term with a shell drawn from an actual profile,
we select $r/L=0.49$ in the underdensity model of
Fig.~\ref{fig:density-profiles}.  This representative shell is marked by the
dotted line in the upper panel and has, at the present epoch,
\begin{equation}
r/L=0.49,\qquad \Delta=-0.337,\qquad\delta=0.103,
\label{eq:representative-shell}
\end{equation}
so it is locally overdense while still enclosing a mass deficit.  We hold
these two contrasts fixed across the following scan to isolate the response
closure from changes in the shell profile.
We compare flat backgrounds with $\Omega_{m0}=0.2,0.3,0.4$, curved backgrounds
with $(\Omega_{m0},\Omega_{k0})=(0.3,\pm0.1)$, and the flat
$\Omega_{m0}=0.3$ model at $z=1,2,5$.  Every closure uses the exact growth rate
of the corresponding background, so the residual isolates the approximation
to $A$ and $A'$.

Table~\ref{tab:backend-errors} shows the result.  The linear response already
captures the qualitative ordering of the rates, but its error is appreciable
for the nonlinear density pair in Eq.~\eqref{eq:representative-shell}.  The
second-order expansion captures most of the leading nonlinear correction
without an equal-age root, reducing the four maximum errors by factors of
$3.5$--$6.5$ relative to linear theory.  M26 reduces the maximum relative error
by factors of roughly $23$--$29$ relative to the linear result and improves on
M13 for all four observables in this test.  The relative error in $\delta_{h_{\rm loc}}$
is numerically amplified because the exact $\delta_{h_{\rm loc}}$ is small; the
corresponding maximum absolute error is $3.0\times10^{-4}$.

\subsection{Exact reconstruction of smooth LTB profiles}
\label{sec:profile-validation}

The decisive test is against a complete spherical spacetime rather than
isolated top hats.  We use the compensated $\LLTB$ underdensity and overdensity
profiles with central present-day contrasts $\delta_0=-0.6$ and $+0.6$ shown in Figure~\ref{fig:density-profiles}.  The exact LTB
calculation supplies $R$, $R'$, $\dot R$, and $\dot R'$ and hence the rates in
Eq.~\eqref{eq:directional-H}.  Independently, we retain only the resulting
$\delta(t,r)$ and $\Delta(t,r)$ and apply the response equations.

\begin{figure*}
\includegraphics[width=\textwidth]{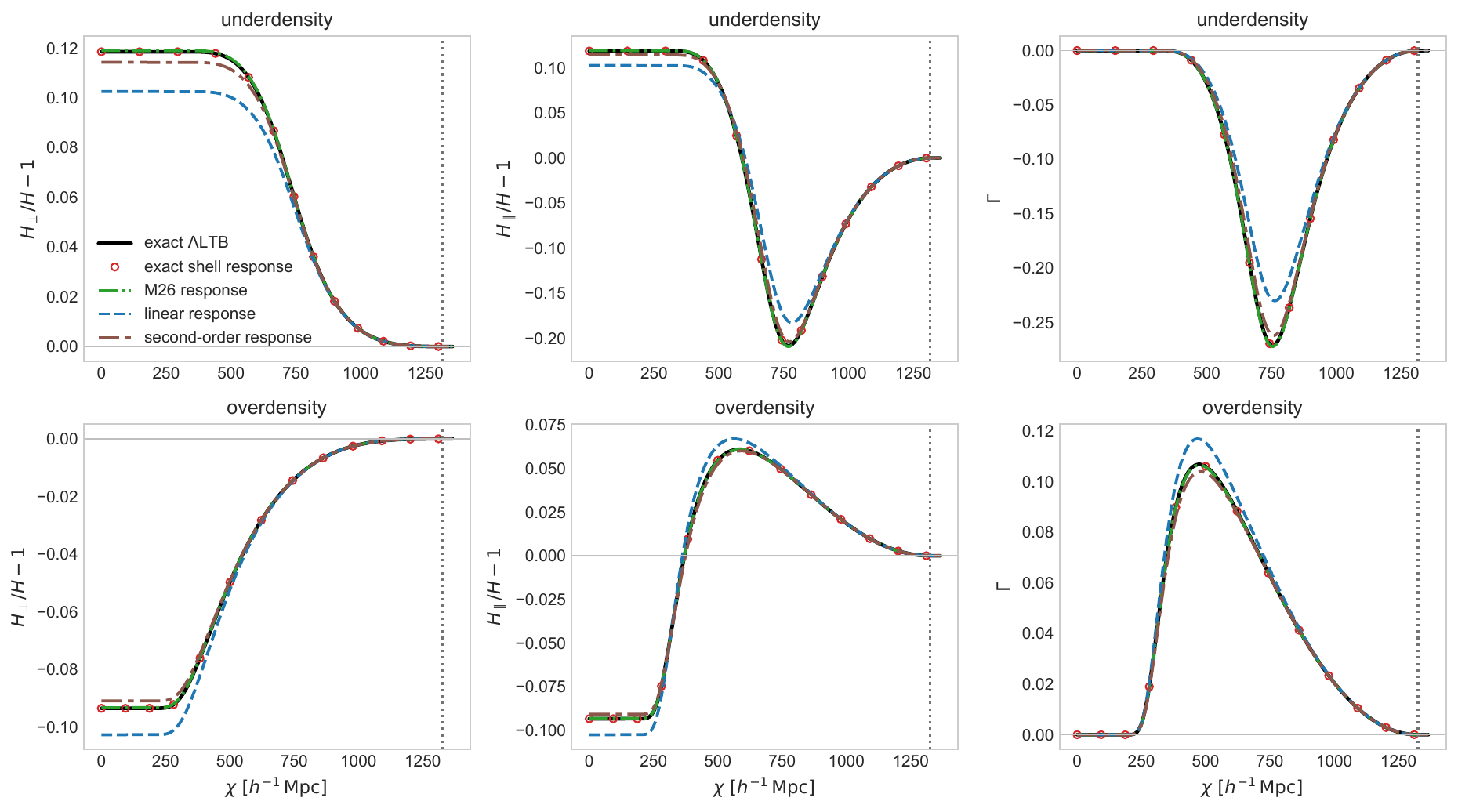}
\caption{Exact present-day LTB expansion contrasts compared with the
root-free linear, second-order, and M26 responses. The exact shell closure is
overlaid as sparse open markers, and vertical dotted lines mark the matching
radii. See Fig.~\ref{fig:density-profiles} for the corresponding density
profiles.}
\label{fig:profile-approximations}
\end{figure*}

For the exact equal-age closure, the two calculations agree at every sampled
radius at floating-point accuracy.  Thus, under
the stated assumptions, the response is not an approximation to LTB dynamics;
it is an exact rewriting of it.
Figure~\ref{fig:profile-approximations} displays this identity.  The transverse response
is controlled by $\Delta$, whereas the radial rate and $\Gamma$ respond most
strongly in the compensated transition, where $\delta-\Delta$ is largest, as shown in Figure~\ref{fig:density-profiles}.

The fast closures can be tested on the same profiles.  Their errors are given
in Table~\ref{tab:profile-errors} and their radial dependence is shown in
Fig.~\ref{fig:profile-approximations}.  The second-order expansion captures
the leading departure from the linear curves, while M26 reconstructs the
profiles to errors of order $10^{-3}$ or smaller.  This very good reconstruction is especially useful because
the input remains only $f,\delta,\Delta$.

\begin{table}
\caption{Maximum absolute errors on the present-day LTB profiles.  Here
$\mathrm{err}(X)\equiv |X_{\rm approx}-X_{\rm exact}|$.  Errors in the expansion
contrasts are therefore normalized by the background~$H$, whereas the
$\Gamma$ error is absolute.}
\label{tab:profile-errors}
\begin{ruledtabular}
\begin{tabular}{llccc}
profile & closure & $\mathrm{err}(\delta_{h_\perp})$ &
$\mathrm{err}(\delta_{h_\parallel})$ & $\mathrm{err}(\Gamma)$ \\
\hline
underdensity & linear & $1.6\!\times\!10^{-2}$ & $3.5\!\times\!10^{-2}$ & $4.3\!\times\!10^{-2}$ \\
          & second order & $4.3\!\times\!10^{-3}$ & $1.0\!\times\!10^{-2}$ & $1.2\!\times\!10^{-2}$ \\
          & M26 & $4.4\!\times\!10^{-4}$ & $1.0\!\times\!10^{-3}$ & $1.3\!\times\!10^{-3}$ \\
overdensity  & linear & $9.2\!\times\!10^{-3}$ & $9.3\!\times\!10^{-3}$ & $1.0\!\times\!10^{-2}$ \\
          & second order & $2.5\!\times\!10^{-3}$ & $3.2\!\times\!10^{-3}$ & $4.0\!\times\!10^{-3}$ \\
          & M26 & $2.0\!\times\!10^{-4}$ & $2.0\!\times\!10^{-4}$ & $3.1\!\times\!10^{-4}$ \\
\end{tabular}
\end{ruledtabular}
\end{table}

\section{Discussion and conclusions}
\label{sec:discussion}

We have reduced the growing-mode density--expansion relation for every shell
of a smooth spherical profile to one top-hat response and its derivative.  The
enclosed contrast $\Delta$ fixes the transverse rate through
$A(\Delta)=\Delta\Theta_\perp(\Delta)$, while the difference
$\delta-\Delta$ and the derivative $A'$ supply all additional information
needed for $\Hpar$, $\Hloc$, and $\Gamma$.  The linear limit is therefore an
algebraic map from $f,\delta,\Delta$, and its EdS-kernel second-order
extension captures most of the leading nonlinear correction without a root.
Nonlinearly, the equal-age
construction gives the exact response and reproduces complete growing-mode
$\LLTB$ profiles at floating-point precision.

The M26 fit provides a nearly universal, root-free approximation to the nonlinear correction.  At
the representative gradient shell of Section~\ref{sec:across-bkg}, its maximum
relative errors across the tested backgrounds are $0.3\%$, $0.7\%$, and
$0.6\%$ for $\delta_{h_\perp}$, $\delta_{h_\parallel}$, and $\Gamma$,
respectively.  On the complete profiles, the absolute errors in the
directional expansion contrasts are of order $10^{-3}$ or smaller.
Comparisons with B92, BC08, and NG13
also show why fitting $A'$ matters: a relation can reproduce the top-hat
response accurately while being less accurate for quantities controlled by a
radial derivative.

The weak cosmology dependence of the nonlinear correction has a simple
interpretation.  Changing the background cosmology mainly changes the rate at
which structure evolves, rather than the path followed by the density and
velocity fields.  The linear growth factor $D$ and growth rate
$f=\dd\ln D/\dd\ln a$ absorb most of this change of clock.  Nusser and Colberg
made this explicit by using $\tau=\ln D$ as time and scaling the velocity by
$Hf$ \cite{Nusser:1997ec}.  In these variables the continuity equation loses
its explicit cosmology dependence, while the leading residual in the Euler
equation is proportional to
$\epsilon=\Omega_m/f^2-1$.  This residual vanishes in Einstein--de Sitter and
does not affect the linear growing mode; it enters only through nonlinear
departures between the rescaled velocity and gravitational fields.  The
remaining cosmology dependence is consequently small and most visible at late
times.

Perturbation theory and the exact shell construction express the same idea in
complementary ways.  Higher-order growth functions remain close to their
Einstein--de Sitter values after the dominant powers of $D$ and $f$ are
removed~\cite{Fosalba:1998da}, and phase-space treatments find only a weak
residual dependence after the growth rate is factored out~\cite{Nadkarni-Ghosh:2012hmw}.  In the equal-age construction, changing the
cosmology shifts the ages of both the shell and the reference background in a
similar way; division by $f$ removes most of the resulting change in their
relative expansion.  A small physical width remains because the shell
curvature required to produce a fixed instantaneous $\Delta$ depends on the
background age and growth history.  At fixed density parameters, $H_0$ cancels
exactly from the dimensionless response.  M26 therefore represents the center
of a narrow family of response curves, rather than an exactly universal law.
The same near universality is seen in smoothed $N$-body density--velocity
fields and in joint statistics of concentric spheres
\cite{Nusser:1997ec,Uhlemann:2016wug}; it should not, however, be extrapolated
to virialized halo interiors or other post-shell-crossing dynamics.

The extension from a top hat to a smooth profile is essential here.  A top hat
sets $\delta=\Delta$ and hides $A'$.  In a general profile, $A'$ converts the
local-minus-enclosed density contrast into the radial response.  Consequently,
$\Gamma$ vanishes where a shell has the same density as its interior average
and becomes largest across strong radial gradients.  It is therefore a
diagnostic of inhomogeneous background structure, not of stochastic
small-scale perturbations.  This is also why the calibration in
Appendix~\ref{app:calibration} constrains both $\Theta_\perp$ and $A'$.

The exact nonlinear closure assumes pressureless matter, a homogeneous
cosmological constant, a simultaneous bang time, spherical symmetry, and no
shell crossing.  It describes the expanding branch up to turnaround, but does
not follow the subsequent collapsing and virializing evolution, nor does it
include radiation, pressure support, baryonic physics, or nonspherical
collapse.  In a triaxial system the density alone is insufficient: the
deformation, velocity-gradient, and tidal tensors must also be evolved
\cite{Nadkarni-Ghosh:2014bwa}.  The linear relations are less restrictive
with respect to the background cosmology, but not with respect to geometry.
Within spherical symmetry,
Eqs.~\eqref{eq:linear-perp}--\eqref{eq:linear-gamma} apply whenever the matter
perturbations have scale-independent linear growth, including smooth $w$CDM,
after supplying the corresponding $H(a)$ and $f(a)$.  The second-order
Eqs.~\eqref{eq:second-order-perp}--\eqref{eq:second-order-gamma} additionally
adopt the EdS kernels of Eqs.~\eqref{eq:eds-nu2} and \eqref{eq:eds-c2}; away
from EdS they are an approximation justified by the weak cosmology dependence
of the higher-order growth functions, not an exact general-$w$ result.
By contrast, M26 is a nonlinear spherical approximation calibrated for
$w=-1$. Existing spherical-collapse results suggest that, after growth-rate
normalization, the residual dependence on $w$ is weak
\cite{Nadkarni-Ghosh:2012hmw}; nevertheless, M26 has not been validated away
from $w=-1$ and should be used within the domain specified in
Eq.~\eqref{eq:calibration-domain}.

The formalism deliberately does not evolve the density profile.  Instead, the
separation in Eq.~\eqref{eq:conceptual-map} allows a spherical model,
simulation, or reconstruction to provide $\delta,\Delta$, after which the
directional expansion follows without re-solving the radial dynamics or
differentiating a metric table.  Conversely, directional expansion
measurements could constrain the local and enclosed density pair, although
applications to data will require lightcone observables and realistic
selection functions.
The main conclusion is that one familiar top-hat response contains
the complete directional expansion information for any growing, smooth
spherical profile before shell crossing, provided that its ordinary derivative
is retained.  This separation of profile evolution from expansion response
makes the dynamics easier to interpret, validate, and reuse.

\begin{acknowledgments}
VM thanks CNPq (Brazil), CAPES (Brazil) and FAPES (Brazil) for partial financial support.
\end{acknowledgments}

\appendix

\section{M26 fit}
\label{app:calibration}

We search transparent expressions of the form
\begin{equation}
\Theta(\Delta)=1+\sum_{i=1}^{N}c_i g_i(\Delta),
\qquad N\le3,
\label{eq:fit-family}
\end{equation}
where every primitive satisfies $g_i(0)=0$.  Consequently every candidate
obeys $\Theta(0)=A'(0)=1$ exactly, with
\begin{equation}
A'(\Delta)=1+\sum_i c_i
\left[g_i(\Delta)+\Delta g_i'(\Delta)\right].
\label{eq:fit-derivative}
\end{equation}
The primitive library contains low-order powers of $\Delta$,
$\ln(1+\Delta)$ and its low powers, simple fractional powers of
$1+\Delta$, $\sin\Delta$, $\tanh\Delta$, and a small set of domain-safe
rational terms.  We enumerate every one-, two-, and three-term combination.
For a fixed combination the coefficients enter linearly, so the
least-squares solution is deterministic and no nonlinear parameter search is
required.

Exact targets were generated on 157 contrast values over
$-0.9\le\Delta\le3$, including $\Delta=0$ exactly.  Complete background
cosmologies, rather than individual contrast samples, were assigned to the
training and validation sets.  The training grid was
\begin{align}
\Omega_{m0} \!\in\!\{0.2,0.3,0.4\}, \,
\Omega_{k0}\!\in\!\{-0.1,0,0.1\}, \,
z\!\in\!\{0,1,5\},
\label{eq:training-grid}
\end{align}
giving 27 backgrounds.  The held-out interpolation grid was
\begin{align}
\Omega_{m0}\!\in\!\{0.25,0.35\}, \,
\Omega_{k0}\!\in\!\{-0.05,0.05\}, \,
z\!\in\!\{0.5,2\},
\label{eq:validation-grid}
\end{align}
giving eight previously unseen backgrounds.

For a training set $S$, the minimized loss was
\begin{align}
{\cal L}_\lambda \!=\!
\left\langle \! \!
\left(\frac{\Theta_{\rm fit}\!-\!\Theta_{\rm exact}}{s_\Theta}\!\right)^2
\right\rangle_S
\!\!\!\!+\!\!\lambda
\left\langle \!\!
\left(\frac{A'_{\rm fit}\!-\!A'_{\rm exact}}{s_{A'}}\!\right)^2
\right\rangle_S \!\!\!.
\label{eq:fit-loss}
\end{align}
The fixed scales $s_\Theta=2.83\times10^{-3}$ and
$s_{A'}=1.11\times10^{-2}$ are the RMS errors, on the same extended training
set, of a preliminary refit within the M13 three-parameter family
$1-a\Delta-b\sin\Delta/(c+\Delta)$, used only as a reference.  To select both
$\lambda$ and the symbolic expression, let $E_{q,p}$ be a candidate's training
error for $q\in\{\Theta,A'\}$ and $p\in\{\mathrm{RMS},\max\}$.  We minimized
\begin{equation}
{\cal B}=\frac{1}{4}
\sum_{q\in\{\Theta,A'\}}
\sum_{p\in\{\mathrm{RMS},\max\}}
\frac{E_{q,p}}{E^{\rm ref}_{q,p}},
\label{eq:fit-balanced-score}
\end{equation}
where $E^{\rm ref}_{q,p}$ is the corresponding preliminary-fit error.  In the
order $(\Theta_{\rm RMS},\Theta_{\max},A'_{\rm RMS},A'_{\max})$, these four
denominators are $(2.83\times10^{-3},1.46\times10^{-2},
1.11\times10^{-2},6.13\times10^{-2})$.
Scanning $\lambda\in\{0.1,0.3,1,3,10,30\}$ selected $\lambda=0.1$ and the
primitives $\{\Delta,\sqrt{1+\Delta}-1,\sin\Delta\}$.  The validation
cosmologies were not used in this selection.  On the held-out grid, the rounded
M26 relation gives $(\mathrm{RMS},\max)=(1.45\times10^{-3},7.82\times10^{-3})$
for $\Theta$ and $(5.59\times10^{-3},4.31\times10^{-2})$ for $A'$.

\section{{Covariant interpretation and \texorpdfstring{$q$}{q}-scalar dictionary}}
\label{app:covariant}

{This appendix connects the density--expansion variables used in the
main text to the established covariant and $q$-scalar descriptions of
$\LLTB$ models
\cite{Sussman:2008wx,Sussman:2012xc,Sussman:2013yq,Sussman:2014wua}.
The identities and evolution equations below belong to that formalism.
We retain $c=1$ and display Newton's constant $G$ explicitly.
The symbol $\Lambda$ denotes the cosmological constant in Einstein's
equations, so its equivalent vacuum energy density is
$\rho_\Lambda=\Lambda/(8\pi G)$.}

\subsection{{Dust expansion and shear}}
\label{app:covariant-kinematics}

{Following Sussman's covariant formulation
\cite[Sec.~2.1 and Appendix~D]{Sussman:2012xc}, let $u^a$ be the unit dust four-velocity and
$h_{ab}=g_{ab}+u_a u_b$ the rest-space projector.
The expansion tensor, its trace, and the shear are}
\begin{equation}
{
\begin{aligned}
\Theta_{ab}&=h_a{}^c h_b{}^d\nabla_{(c}u_{d)},\\
\Theta&=h^{ab}\Theta_{ab}=\nabla_a u^a,\qquad
\sigma_{ab}&=\Theta_{ab}-\frac{\Theta}{3}h_{ab}.
\end{aligned}}
\label{eq:covariant-expansion}
\end{equation}
{With $n^a$ the radial unit vector orthogonal to $u^a$, spherical
symmetry gives}
\begin{equation}
{
\Theta_{ab}=\Hpar n_a n_b+\Hperp(h_{ab}-n_a n_b).
}
\label{eq:covariant-eigenvalues}
\end{equation}
{Thus $\Hpar$ and $\Hperp$ are expansion eigenvalues.
In coordinates comoving with the dust they reduce to
Eq.~\eqref{eq:directional-H}.  Their covariant meaning is tied to the
physical dust congruence $u^a$, which specifies the matter flow.
Writing $\sigma_{ab}=\Sigma \, e_{ab}$, with
$e_{ab}=h_{ab}-3n_a n_b$, fixes the shear convention as}
\begin{equation}
{
\begin{aligned}
\Theta&=2\Hperp+\Hpar,\qquad
\Sigma=\frac{\Hperp-\Hpar}{3},\\
\Hperp&=\frac{\Theta}{3}+\Sigma,\qquad
\Hpar=\frac{\Theta}{3}-2\Sigma.
\end{aligned}}
\label{eq:covariant-shear-dictionary}
\end{equation}
{The eigenvalues of $e^a{}_b$ in an orthonormal dust frame are
$(0,-2,1,1)$, with the radial entry second.  Consequently}
\begin{equation}
{
\Hloc=\frac{\Theta}{3},\qquad
\Gamma=-\frac{3\Sigma}{\Hloc},\qquad
\frac12\sigma_{ab}\sigma^{ab}=3\Sigma^2.
}
\label{eq:covariant-gamma}
\end{equation}
{The signed anisotropy $\Gamma$ therefore measures the shear relative
to the local volume expansion, wherever $\Hloc\ne0$.}

\subsection{{Density contrasts, averaging, and Weyl curvature}}
\label{app:q-dictionary}

{For this appendix, define
$\rho(t,r)\equiv\rho_m(t,r)$, $\rho_b(t)\equiv\rho_m(t)$, and
$H_b(t)\equiv H(t)$, where $b$ stands for the background.
For a regular spherical domain extending from the center to shell $r$,
the $q$-average of a scalar $F$ is
\cite[Eq.~(13)]{Sussman:2012xc}}
\begin{equation}
{
\begin{aligned}
F_q(t,r)\equiv\langle F\rangle_q(t,r)
\!=\!
\frac{\displaystyle\int_0^r \!\! F(t,\bar r)R^2(t,\bar r)R'(t,\bar r)\,\dd\bar r}
{\displaystyle\int_0^r
 R^2(t,\bar r)R'(t,\bar r)\,\dd\bar r}.
\end{aligned}
}
\label{eq:q-average-definition}
\end{equation}
{For the basic scalars $F\in\{\rho,\Hloc,K\}$,
$K\equiv{}^{(3)}\!{\cal R}/6$ is the intrinsic scalar curvature of the dust
rest spaces divided by six.  The denominator is $R^3(t,r)/3$.
The proper-volume element is
$\dd V_p=4\pi R^2R'\dd r/\sqrt{1-k(r)r^2}$.
Hence Eq.~\eqref{eq:q-average-definition} is a proper-volume average
weighted by $\sqrt{1-k(r)r^2}$; its measure is precisely the
Euclidean-volume measure used for the enclosed contrast in the main text
\cite{Sussman:2012xc,Sussman:2013yq}.  The resulting $q$-scalars are
\cite[Eq.~(17)]{Sussman:2012xc}}
\begin{equation}
{
\begin{aligned}
\rho_q&=\frac{3M}{4\pi R^3},\quad H_q=\frac{\dot R}{R}=\Hperp, \quad
K_q&=\frac{k(r)r^2}{R^2}.
\end{aligned}
}
\label{eq:q-scalars-dictionary}
\end{equation}
{Here $M$ is the conserved dust contribution to the Misner--Sharp
mass; the homogeneous vacuum contribution is kept explicit as $\Lambda$.
In particular,}
\begin{equation}
{
\begin{aligned}
\delta=\frac{\rho}{\rho_b}-1,\quad
\Delta=\frac{\rho_q}{\rho_b}-1=\langle\delta\rangle_q, \quad
\delta_q^{(\rho)}=\frac{\delta-\Delta}{1+\Delta}.
\end{aligned}
}
\label{eq:q-contrasts-dictionary}
\end{equation}
{The relative $q$-fluctuation $\delta_q^{(\rho)}$ compares the local
density with the enclosed density, whereas $\delta$ and $\Delta$ both use
the same FLRW reference \cite[Sec.~5]{Sussman:2013yq}.
Define also the absolute fluctuations}
\begin{equation}
{
\begin{aligned}
{\cal D}^{\rho}&=\rho-\rho_q=\rho_b(\delta-\Delta),\\
{\cal D}^{H}&=\Hloc-H_q=-\Sigma,\qquad
{\cal D}^{K}=K-K_q.
\end{aligned}}
\label{eq:q-absolute-fluctuations}
\end{equation}
{Differentiating Eq.~\eqref{eq:q-average-definition} gives the exact
identity \cite[Eq.~(21)]{Sussman:2012xc}}
\begin{equation}
{
F-F_q=\frac{R}{3R'}\partial_r F_q.
}
\label{eq:q-gradient-dictionary}
\end{equation}
{For density and expansion this yields
Eqs.~\eqref{eq:Delta-radial-identity} and
\eqref{eq:H-radial-identity}, respectively.}

{The electric Weyl tensor
$\mathcal{E}_{ab}=C_{acbd}u^c u^d$ has the same spherical eigenframe as the shear:
$\mathcal{E}_{ab}=\mathcal{E}e_{ab}$, with the convention
\cite[Eq.~(37)]{Sussman:2012xc}}
\begin{equation}
{
{\cal E}=-\frac{4\pi G}{3}(\rho-\rho_q)
=\frac{4\pi G\rho_b}{3}(\Delta-\delta).
}
\label{eq:q-weyl-dictionary}
\end{equation}
{Thus the local-to-enclosed density difference measures the electric
Weyl curvature, while the directional expansion difference measures shear.
For the common-age growing-mode response in the main text,}
\begin{equation}
{
\begin{aligned}
\Sigma =\frac{fH_b}{3}(\delta-\Delta)A'(\Delta)
=-\frac{fH_b}{4\pi G\rho_b}A'(\Delta)\,{\cal E}.
\end{aligned}}
\label{eq:response-shear-weyl}
\end{equation}
{This is the covariant interpretation of the derivative term in
Eq.~\eqref{eq:Hparallel-response}: the response relates the density
inhomogeneity encoded in Weyl curvature to anisotropic dust expansion.}

\subsection{{Evolution equations and constraints}}
\label{app:covariant-evolution}

{The corresponding standard dust evolution equations can be written
in these variables without imposing the common-age restriction
\cite[Eqs.~(8)--(11)]{Sussman:2014wua}.  A dot denotes the derivative along the dust flow,
and the reference FLRW epoch is identified with the same proper time.
The background equations are}
\begin{align}
{\dot\rho_b}&{=-3H_b\rho_b,}
\label{eq:q-background-density}\\
{\dot H_b}&{=-H_b^2-\frac{4\pi G}{3}\rho_b+\frac{\Lambda}{3}.}
\label{eq:q-background-expansion}
\end{align}
{Local and enclosed dust conservation give}
\begin{align}
{\dot\delta}&{=-3(1+\delta)(H_q+{\cal D}^{H}-H_b),}
\label{eq:q-local-contrast-evolution}\\
{\dot\Delta}&{=-3(1+\Delta)(H_q-H_b).}
\label{eq:q-enclosed-contrast-evolution}
\end{align}
{The expansion variables obey}
\begin{align}
{\dot H_q}&{=-H_q^2-\frac{4\pi G}{3}\rho_b(1+\Delta)
+\frac{\Lambda}{3},}
\label{eq:q-expansion-evolution}\\
{\dot{\cal D}^{H}}&{=-(2H_q+3{\cal D}^{H}){\cal D}^{H}
-\frac{4\pi G}{3}\rho_b(\delta-\Delta).}
\label{eq:q-expansion-fluctuation-evolution}
\end{align}
{The algebraic constraints are}
\begin{align}
{H_q^2}&{=\frac{8\pi G}{3}\rho_b(1+\Delta)-K_q
+\frac{\Lambda}{3},}
\label{eq:q-energy-constraint}\\
{2H_q{\cal D}^{H}}&{=\frac{8\pi G}{3}\rho_b(\delta-\Delta)
-{\cal D}^{K}.}
\label{eq:q-fluctuation-constraint}
\end{align}
{Together with Eq.~\eqref{eq:q-gradient-dictionary}, these constrain
the radial initial data.  They also recover the local energy constraint
$\Hloc^2=8\pi G\rho/3+\Lambda/3-K+\Sigma^2$.}

{If metric histories are required, two further kinematic equations
complete the description:}
\begin{align}
{\dot a_\perp}&{=H_q a_\perp, \quad
\dot{\cal J}}&{=3{\cal D}^{H}{\cal J},\quad
{\cal J}\equiv\frac{a_\parallel}{a_\perp}=\frac{rR'}{R}.}
\label{eq:q-radial-stretch-evolution}
\end{align}
{The radial stretch $\mathcal{J}$ is distinct from the expansion
anisotropy $\Gamma$ of Eq.~\eqref{eq:Hloc-Gamma}.  At each shell the
evolution is an ordinary differential system in proper time, with radial
dependence supplied by initial data satisfying the constraints.
In the FLRW limit $H_q=H_b$, $\mathcal{D}^{H}=0$, and
$\delta=\Delta=0$, both contrast derivatives vanish.
The general evolution system admits the usual growing and decaying modes;
the equal-age construction in Sec.~\ref{sec:response} selects the
simultaneous-bang, expanding solution used for our response.
These equations connect the response variables to the full dust dynamics,
while the calculation in the main text reconstructs the instantaneous
expansion from $\delta,\Delta$ without integrating this system.}

\bibliography{references}

\end{document}